\documentclass[twocolumn]{aastex701}
\usepackage{CJK}
\usepackage{url} % 提供\url{}命令，支持长字符换行
\usepackage{seqsplit} % 提供\seqsplit{}命令，强制拆分长字符
\usepackage{xcolor}  % 颜色基础包（必加）

\usepackage{booktabs} % For professional table lines
\usepackage{multirow} % For merged cells

\newcommand{\BT}{B\slash T}

\newcommand{\eb}{e_\mathrm{Bar}}
\newcommand{\fb}{f_\mathrm{Bar}}

\newcommand{\logm}{\log M_\star}
\newcommand{\logmr}{\log (M_\star / M_{\mathrm{neigh}})}

\newcommand{\kpc}{\ h^{-1}\ \text{kpc}}
\newcommand{\kms}{\ \text{km}\ \text{s}^{-1}}
\newcommand{\pdis}{d_{\text{p}}}
\newcommand{\dv}{\Delta v}

\begin{document}
\begin{CJK*}{UTF8}{gbsn}

\received{\today}
\revised{\today}
\accepted{\today}

\submitjournal{ApJ}

\shorttitle{Bar Fraction of Galaxy Pairs}
\shortauthors{Li et al.}

\title{Reduction of bar fraction in paired galaxies in the SDSS}

\correspondingauthor{Shuai Feng}
\email{sfeng@hebtu.edu.cn}

\author[0000-0003-1454-2268]{Linlin Li (李林林)}
\affiliation{College of Physics, Hebei Normal University, 20 South Erhuan Road, Shijiazhuang 050024, China}
\affiliation{Shijiazhuang Key Laboratory of Astronomy and Space Science / Guoshoujing Institute of Astronomy, \\
Hebei Normal University, Shijiazhuang, Hebei 050024, China}
\email{lilin@hebtu.edu.cn}

\author[0000-0002-9767-9237]{Shuai Feng (冯帅)}
\affiliation{College of Physics, Hebei Normal University, 20 South Erhuan Road, Shijiazhuang 050024, China}
\affiliation{Shijiazhuang Key Laboratory of Astronomy and Space Science / Guoshoujing Institute of Astronomy, \\
Hebei Normal University, Shijiazhuang, Hebei 050024, China}
\affiliation{Hebei Key Laboratory of Photophysics Research and Application, Shijiazhuang 050024, China}
\email{sfeng@hebtu.edu.cn}

\author[0000-0002-3073-5871]{Shiyin Shen (沈世银)}
\affiliation{Shanghai Astronomical Observatory, Chinese Academy of Sciences, 80 Nandan Road, Shanghai 200030, China}
\affiliation{Shanghai Key Lab for Astrophysics, Shanghai 200234, China\\}
\email{ssy@shao.ac.cn}

\author[0009-0006-5550-4225]{Qi'an Deng (邓淇安)}
\affiliation{Department of Astronomy, School of Physics and Astronomy, Shanghai Jiao Tong University, Shanghai 200225, China}
\email{kennydeng@sjtu.edu.cn}

\author[0000-0001-6966-6925]{Ying Zu (祖颖)}
\affiliation{State Key Laboratory of Dark Matter Physics, Tsung-Dao Lee Institute \& School of Physics and Astronomy, Shanghai Jiao Tong University, Shanghai, 200240, China}
\email{yingzu@sjtu.edu.cn}

\author[0000-0003-1259-9908]{Wenyuan Cui (崔文元)}
\affiliation{College of Physics, Hebei Normal University, 20 South Erhuan Road, Shijiazhuang 050024, China}
\affiliation{Shijiazhuang Key Laboratory of Astronomy and Space Science / Guoshoujing Institute of Astronomy, \\
Hebei Normal University, Shijiazhuang, Hebei 050024, China}
\email{wenyuancui@126.com}

\begin{abstract}
We investigate the bar fraction in galaxy pairs from the SDSS to assess how galaxy interactions affect bar structures. Compared to isolated galaxies, close pairs exhibit a significantly reduced bar fraction at projected separations within $25\kpc$. This reduction is driven almost entirely by systems showing clear merger or disturbance signatures, indicating that tidal interactions suppress bars. The decline is dominated by a decrease in weak bars, while the fraction of strong bars remains largely unchanged. Bar suppression is primarily associated with major mergers and is strongest in massive host galaxies. A weaker but statistically significant suppression is detected in minor mergers only for massive galaxies with small bulges. In contrast, no significant dependence of bar suppression on the relative orientation between pair members is found. These findings provide observational evidence that tidal perturbations in major mergers play a key role in regulating bar evolution.
\end{abstract}


\section{Introduction}

\label{sec:intro}

Bars are elongated stellar structures commonly found in disk galaxies, playing a crucial role in their internal dynamics and overall evolution. Observations indicate that a significant fraction of disk galaxies host bars, with bar fractions varying depending on observational band, morphological type, and stellar mass 
\citep{Eskridge2000, Nair2010, Masters2011, Erwin2018}. Bars significantly influence the internal kinematics and matter distribution of galaxies, driving angular momentum redistribution and facilitating gas inflow toward the central regions \citep{Sellwood1993, Athanassoula2003, Debattista2004}, which enhances star formation and nuclear activity \citep{Ellison2011, Lin2017, Oh2012}. Given their widespread presence and profound impact on galaxy structure and evolution, understanding the formation and evolution of bars is crucial for studying the broader context of galaxy evolution.

In isolated environments, bars can spontaneously form due to internal dynamical instabilities in the disk \citep{Hohl1971, Sellwood1981}. In dense environments, bars can also form due to external mechanisms such as tidal interactions and galaxy mergers. Tidal force from neighboring galaxies can induce bar structures by perturbing the stellar disk and redistributing angular momentum \citep{Noguchi1987, Miwa1998, MartinezValpuesta2017}. Mergers have been proposed as another possible channel for bar formation by dynamically heating the disk while simultaneously triggering non-axisymmetric instabilities \citep{RomanoDiaz2008, Athanassoula2016, Cavanagh2020}. Beyond direct interactions, environmental effects such as the cluster tidal field and ram pressure stripping can also influence bar formation and evolution \citep{Mastropietro2005, Lokas2016, Lokas2023}. 

Theoretical studies remain divided on how galaxy interactions influence bar formation and evolution. Many simulations suggest that tidal forces can trigger bar formation by amplifying disk instabilities, and even regard interactions as a primary mechanism for bar generation \citep[e.g.,][]{Gerin1990, Berentzen2004}. Further investigations have shown that the outcome depends sensitively on interaction parameters, such as mass ratio and orbital configuration \citep{Lang2014, Cavanagh2020}. Conversely, other simulations find that interactions may suppress bar formation or even weaken and destroy existing bars \citep[e.g.,][]{Lokas2014, Wille2024, Lu2024}. Recent studies further reveal a more complex picture, predicting that bars can be destroyed during the early stages of a merger and subsequently reform later in the interaction \citep{Zana2018, Peschken2019}. These contrasting results highlight the nuanced role of interactions in bar evolution and underscore the importance of observational constraints.

Observational efforts have long explored the relationship between bar structures and galaxy environments as a way to understand the role of interactions in bar formation. However, findings remain mixed. Some studies report that galaxies in denser environments, quantified by parameters such as local galaxy density or distance to the cluster center, tend to have higher bar fractions, suggesting that frequent encounters in these environments may promote bar formation through tidal perturbations \citep[e.g.,][]{Skibba2012, Castignani2022}. In contrast, other works find the opposite trend. The bar fractions decline as the number of neighboring galaxies increases, implying that high-density environments may instead suppress or disrupt bar structures \citep[e.g.,][]{Lee2012, Lin2014}. Still, some studies observe no significant dependence of bar fraction on the environment at all \citep[e.g.,][]{Aguerri2009, Sarkar2021}. These discrepancies suggest that environmental factors influence bars in complex and possibly competing ways, making it difficult to isolate the effect of galaxy interactions.

Importantly, many of these studies rely on large samples of galaxies in groups or clusters, where environmental influences extend beyond direct galaxy-galaxy interactions. In such dense environments, many other factors, like cluster mergers and the global tidal field \citep{Yoon2019, Deng2023}, may contribute to morphological transformations. Galaxy pairs offer a cleaner laboratory to constrain the role of galaxy interactions better. Therefore, a focused study of isolated galaxy pairs is essential to clarify how direct tidal encounters influence bar formation and evolution.

In this work, we conduct a systematic analysis of the bar fraction in galaxy pairs, examining their correlation with other observed galaxy properties. By leveraging a large sample of isolated galaxy pairs, we aim to explore how tidal interactions influence bar formation and evolution while minimizing contamination from broader environmental effects. This paper is organized as follows. In Section \ref{sec:data}, we describe the galaxy pair sample and the bar strength measurements. Section \ref{sec:ressults} presents the dependence of bar fraction on the properties of galaxy pairs. Next, we make relevant discussions in Section \ref{discuss}. Finally, Section \ref{sec:sum} summarizes our findings. Throughout this work, we adopt a standard cosmology with $\Omega_M = 0.3$, $\Omega_{\Lambda} = 0.7$, and $H_0 = 100h\  \text{km}\ \text{s}^{-1}\ \text{Mpc}^{-1}$ with $h = 1$.

\section{Data} \label{sec:data}

\subsection{Galaxy Sample}

The galaxy pair sample is derived from the main galaxy sample of SDSS DR7 \citep{Strauss2002, SDSSDR7, Blanton2005}. To increase the spectral completeness due to the fiber collision effect, we supplement a large number of spectroscopic redshifts from additional spectroscopic surveys \citep[see details in][]{Shen2016, Feng2019}, including LAMOST \citep{Zhao2012, Luo2015} and GAMA \citep{Driver2011, Baldry2018}. Then, we select galaxy pairs with the following criteria: (1) the line-of-sight velocity difference between the two galaxies must satisfy $|\dv| \leq 500 \kms$; (2) the projected separation must be within $10\kpc \leq \pdis \leq 200 \kpc$; and (3) each galaxy can have at most one companion that meets these conditions. Finally, we obtained a sample of $46,634$ isolated galaxy pairs.

In addition to the galaxy pair sample, we also construct an isolated galaxy sample using similar criteria. Specifically, we select galaxies with no companion within a line-of-sight velocity difference of $|\dv| \leq 500 \kms$ and a projected separation of $\pdis \leq 200 \kpc$. We obtained a sample of $529,295$ isolated galaxies. This isolated galaxy sample will be used for selecting a control sample for comparison with galaxy pairs in the following.

\subsection{Bar Strength Measurement}

The bar strength measurements used in this study are directly adopted from \citet{Deng2023}, who developed an automated bar detection method based on ellipse fitting of galaxy isophotes. Their sample consists of low-inclination disk galaxies from the SDSS, selected to have outer isophotal ellipticities (enclosing 90\% of the total light) below $0.5$, stellar masses $\logm > 10$ and bulge-to-total ratios $B/T < 0.5$. Bars were systematically identified by analyzing the ellipticity and position angle (PA) profiles before and after subtracting a two-dimensional disc model. To ensure consistency with our study, we use the subset of their barred galaxy catalog that overlaps with our galaxy pair sample.

The bar strength in \citet{Deng2023} is quantified using the parameter $\eb$, which is a normalized measure of the peak ellipticity detected in the isophote fitting process. This parameter reflects the elongation of the bar structure, with higher values indicating stronger bars. By adopting $\eb$, we ensure a uniform and robust measure of bar strength across our sample. For further details on the measurement process, readers are referred to \citet{Deng2023}. 

For our galaxy pair sample and isolated sample, $20,264$ paired galaxies and $128,689$ isolated galaxies have the bar measurement. 

\subsection{Control Sample}\label{sec:Control-Sample}

To isolate the impact of galaxy interactions on bar structures, we construct a matched control sample of isolated galaxies. Since the bar fraction is known to depend on redshift, stellar mass, morphology, and environment \citep{2004ApJ...612..191E, Masters2011, Nair2010, Skibba2012}, we match each paired galaxy to isolated galaxies with similar redshift ($z$), stellar mass ($\log M_\star$, from \citealt{Kauffmann2003}), bulge-to-total light ratio ($B/T$, from \citealt{Simard2011}), and local environment quantified by the neighbor number density $\Sigma_5$ \citep{Baldry2006}. The matching criteria are
\begin{itemize}
    \item $|\Delta z| < 0.01$,
    \item $|\Delta \log M_\star| < 0.1$,
    \item $|\Delta B/T| < 0.05$,
    \item $|\Delta \log \Sigma_5| < 0.1$.
\end{itemize}
With these controls in place, differences in bar properties between paired and isolated galaxies can be interpreted as the relative impact of galaxy interactions.

\begin{figure*}[ht!]
\plotone{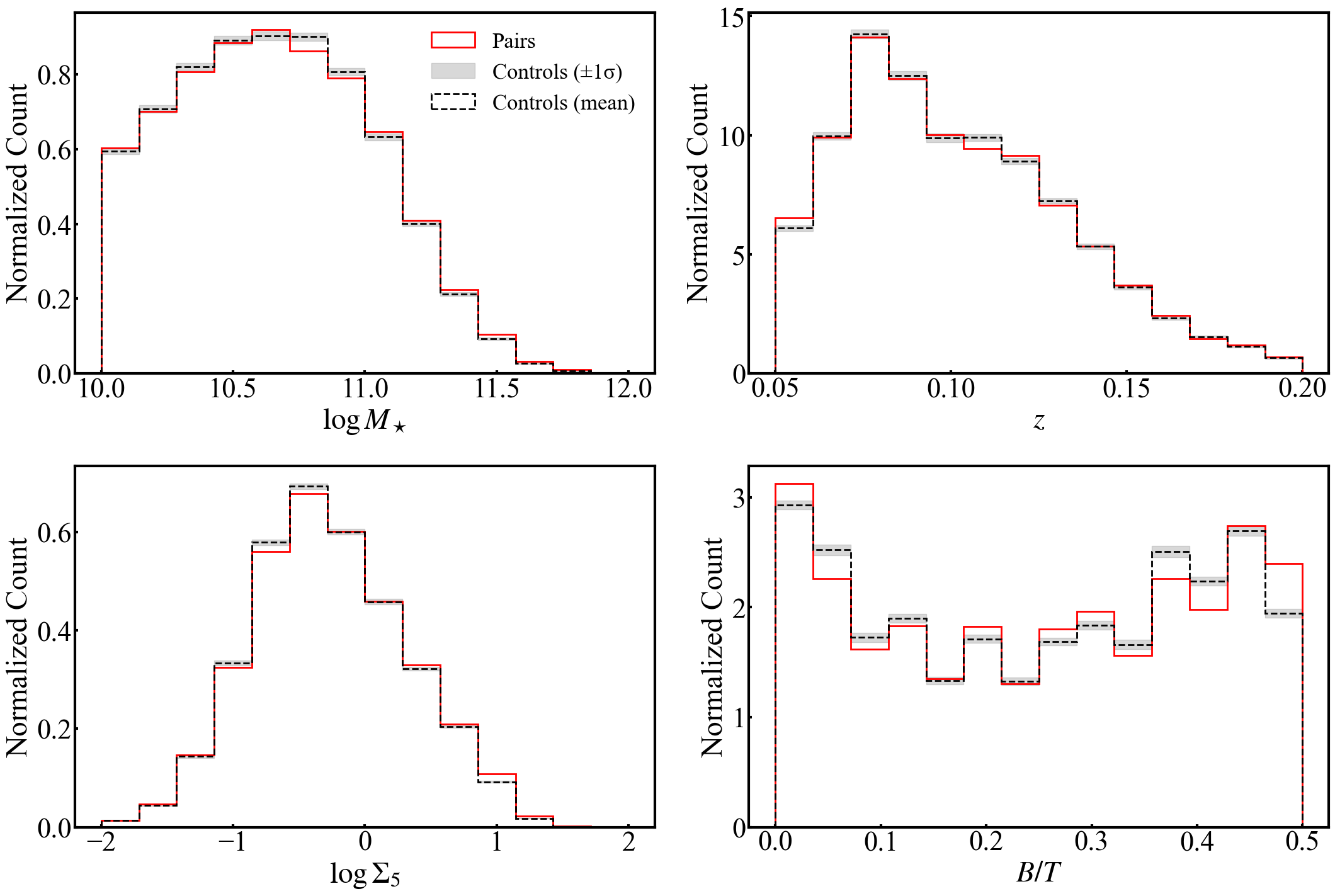}
\caption{Distribution of control parameters for the paired sample and control samples. The red solid lines represent the paired galaxies, while the black dashed lines denote the mean of the $100$ control samples, and the gray shaded regions indicate the $\pm 1\sigma$ scatter across the $100$ iterations. The four panels show the distributions of stellar mass ($\log M_\star$), redshift ($z$), surface density ($\log \Sigma_5$), and bulge-to-total ratio ($B/T$), respectively.}
\label{fig:histsample}
\end{figure*}

For each galaxy in the paired sample, we identify all potential control candidates that satisfy the above criteria and randomly select one candidate per paired galaxy using without-replacement sampling, yielding a $1:1$ matched control sample for each iteration. To account for sampling variability, we repeated this  matching procedure $100$ times to generate $100$ independent control samples. Across these $100$ iterations, the number of paired galaxies successfully matched to at least one eligible isolated control candidate ranges from $20,167$ to $20,181$ (out of the parent $20,264$ paired galaxies), with a mean of $20,173.62$ and a standard deviation of $3.14$, indicating high consistency in the availability of eligible control candidates across iterations. Figure \ref{fig:histsample} presents the distributions of four key control parameters for both the paired sample and the mean of $100$ matched control samples, confirming good statistical matching of the control parameters.

\section{Dependence of Bar Fraction in Galaxy Pairs} \label{sec:ressults}

To assess the impact of galaxy interactions on bar structures, we analyze the barred galaxy fraction $\fb$ as a function of various galaxy pair properties. By comparing $\fb$ between paired galaxies and their matched controls, we aim to isolate the effect of interactions on bar formation and evolution. The barred galaxy fraction $\fb$ is defined as  
\begin{equation}
\fb = \frac{N_\mathrm{Bar}}{N},
\end{equation}  
where $N_\mathrm{Bar}$ is the number of galaxies with a bar strength parameter in the range $0.2 < \eb < 1.0$, and $N$ is the total number of galaxies in each bin. The statistical uncertainties for the fractions are $[ \fb(1-\fb)/N]^{1/2}$ \citep{2008ApJ...675.1141S}.

\subsection{Projected Separation}\label{sec:df-dif}

\begin{figure}[ht!]
    \plotone{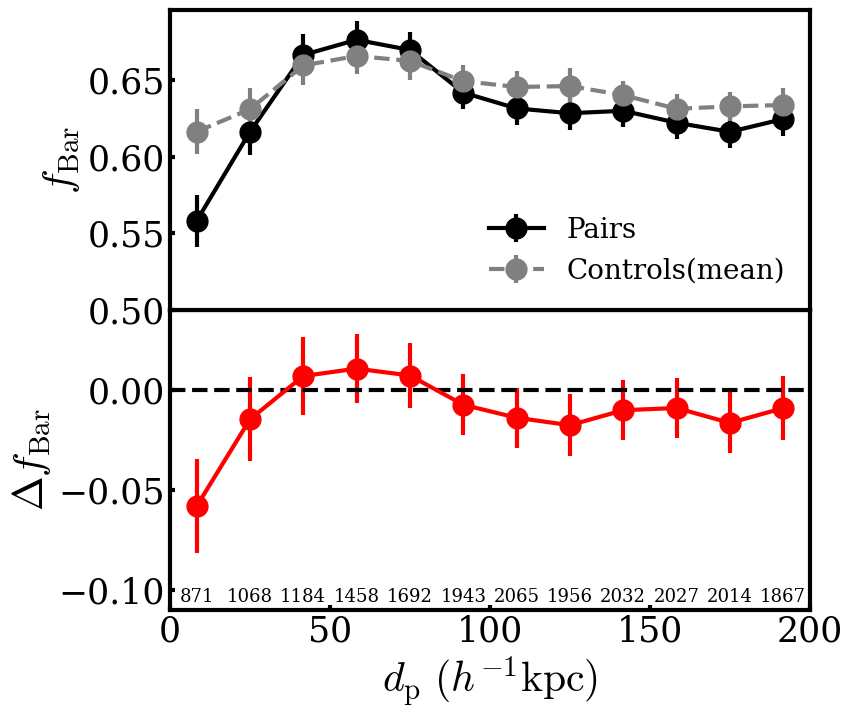}
    \caption{The upper sub-panel shows the overall bar fraction ($\fb$) as a function of projected separation ($\pdis$): black solid circles denote the galaxy pair sample, while gray dashed circles (with error bars) represent the mean of $100$ control samples (error bars indicate the standard error across the $100$ repeats. The lower sub-panel displays the difference in bar fraction ($\Delta \fb$) between galaxy pairs and control samples (red solid circles, with error bars reflecting the combined uncertainty of the pair and control samples). The number of paired galaxies contributing to each data point is indicated above the horizontal axis in the lower sub-panel.}
    \label{fig:df-dis}
\end{figure}

The projected separation between galaxy pairs ($\pdis$) is a commonly used observational proxy for the strength of galaxy interactions. Smaller $\pdis$ values typically correspond to stronger interaction effects. Therefore, we begin our analysis by examining how the barred galaxy fraction ($\fb$) varies with $\pdis$ to assess the influence of interaction strength on bar structures.

The results are presented in the upper panel of Figure \ref{fig:df-dis}\footnote{For figures involving projected separation ($\pdis$) (e.g., Figure \ref{fig:df-dis} and related plots), the $\pdis$ value of each paired galaxy is assigned to its matched isolated control galaxy. This ensures a direct comparison between paired and control samples within the same $\pdis$ bins.}, where the black and gray solid circles show the barred galaxy fraction $\fb$ for paired and control galaxies, respectively. In the control sample, $\fb$ remains nearly constant, ranging from $62\%$ to $68\%$, consistent with the typical bar fraction observed in disk galaxies in the local universe \citep{Eskridge2000, Erwin2018}. In contrast, paired galaxies significantly decrease $\fb$ at small projected separations. To better visualize this effect, we define the difference in bar fraction between paired and control galaxies as,
\begin{equation}
    \Delta\fb = \fb^{\text{pair}} - \fb^{\text{control}},
\end{equation}
which is shown in the lower panel of Figure \ref{fig:df-dis} using red filled circles. To quantify the statistical significance of the $\fb$ differences between paired and control galaxies, we perform two-proportion $Z$-tests in each $\pdis$ bin. We find that only the bin with the smallest projected separation shows a statistically significant difference ($Z = -2.47$, $p = 0.01$), while all other bins show no significant variation ($p > 0.24$). The negative value of $\Delta\fb$ indicates that closely interacting galaxies have a lower bar fraction than their isolated counterparts. To further identify the $\pdis$ range at which the decline becomes statistically significant, we apply a sliding window test. We evaluate the significance of differences in bar fraction within each window using a width of $18\kpc$ and a sliding step of $1\kpc$. The results are insensitive to the exact choice of window width, and we therefore adopt the mean value. This analysis shows a statistically significant reduction in bar fraction at projected separations below $25\kpc$.

This suppression of bar fraction at close separations is consistent with previous studies. For example, \citet{Casteels2013}, using Galaxy Zoo classifications, found a significant decline in the bar fraction within $30\kpc$ of a neighboring galaxy, and \citet{Lin2014} reported reduced clustering of barred galaxies within $50\kpc$ based on SDSS data. These findings suggest that close interactions may disrupt the existing bars. However, in most of these earlier works, the galaxies under consideration often had multiple companions or resided in denser environments. In contrast, our sample specifically selects isolated galaxy pairs, ensuring that each galaxy has only one nearby companion. This cleaner configuration allows for a more direct interpretation of interaction effects. The observed decline in bar fraction at small $\pdis$ therefore provides stronger evidence that close galaxy-galaxy interactions can disrupt existing bar structures.

Above all, our results support a scenario where tidal interactions can suppress or disrupt bar structures in close pairs $\pdis < 25\kpc$. This interpretation is broadly consistent with numerical simulations that show bar structures can be weakened or even destroyed during close passages due to strong tidal forces that disrupt the disk's angular momentum distribution \citep{Lokas2014, Peschken2019, Wille2024}. 

To further understand the origin of this suppression, we now focus on galaxy pairs with $\pdis < 25\kpc$, where the interaction effects are most pronounced. In particular, we examine how the decline in bar fraction correlates with other pair properties, such as the stellar mass of member galaxies and the orientation angle between galaxies, to gain deeper insight into how tidal forces impact bar structures.

\subsection{Stellar Mass of Member Galaxies}\label{sec:df-mmr}
\label{sec:property}

The stellar mass of member galaxies plays a crucial role in determining the strength of interactions within a galaxy pair. More massive galaxies possess deeper gravitational potential wells, making them more resilient to external perturbations, while lower-mass galaxies are more susceptible to tidal effects. Additionally, the mass ratio between paired galaxies can influence interactions. To explore how mass influences bar formation and evolution, we analyze the relationship between the barred galaxy fraction and the stellar mass of galaxy pair members. 

\begin{figure}[ht!]
\plotone{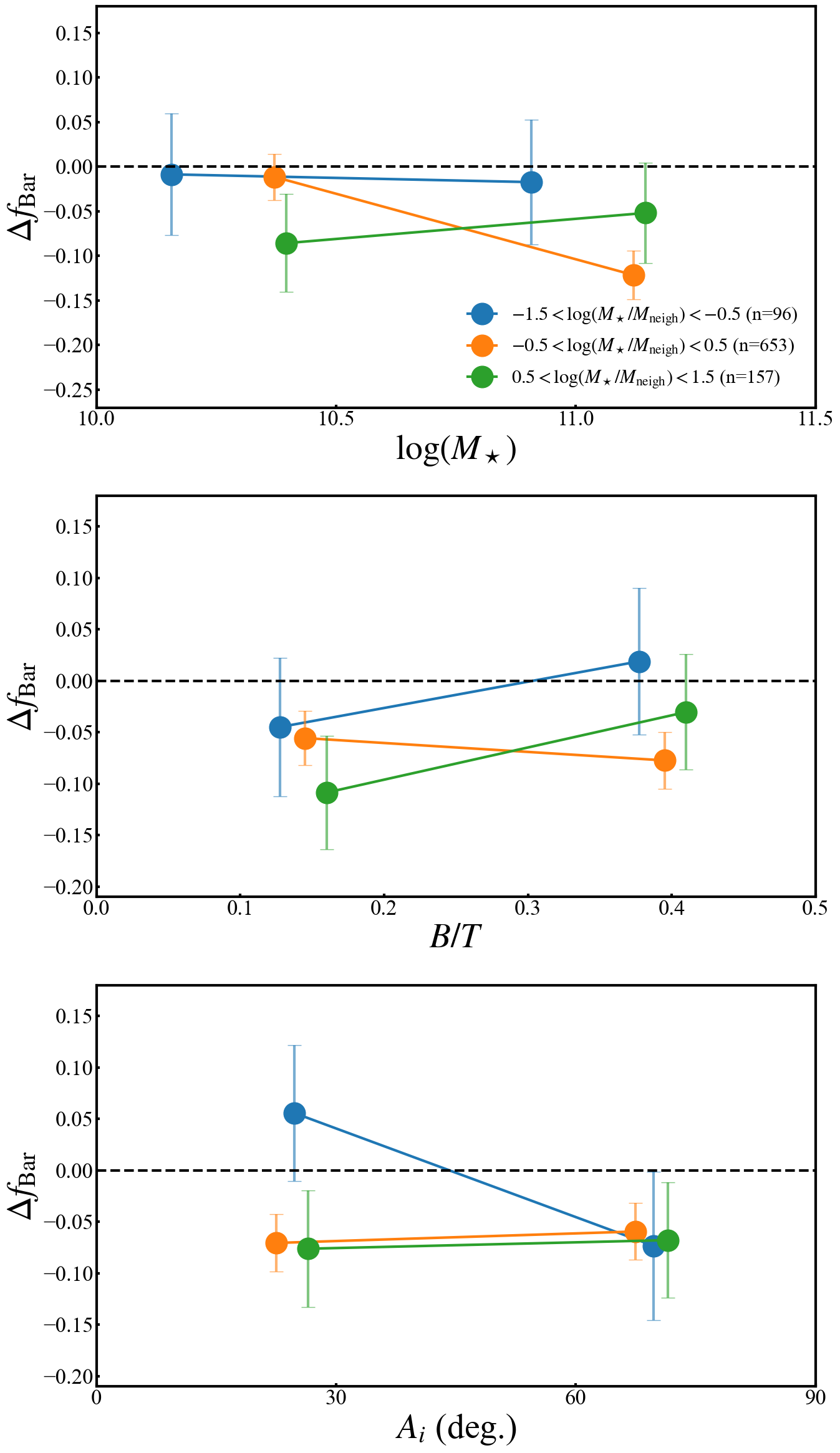}
    \caption{Combined plots of $\Delta\fb$ as a function of galaxy properties for galaxy pairs with projected separations $\pdis < 25\kpc$. \textit{Top}: $\Delta\fb$ versus stellar mass $\logm$; \textit{Middle}: $\Delta\fb$ versus bulge-to-total ratio $\BT$; \textit{Bottom}: $\Delta\fb$ versus relative position angle $A_i$. Colors indicate different mass-ratio intervals, with sample sizes $N$ shown in the legend: blue ($-1.5 < \logmr < -0.5$), orange ($-0.5 < \logmr < 0.5$), and green ($0.5 < \logmr < 1.5$). Error bars represent the uncertainties in $\Delta\fb$. 
\label{fig:df-mmr}}
\end{figure}

The top panel of Figure~\ref{fig:df-mmr} shows $\Delta\fb$ as a function of the stellar mass of the host galaxy \footnote{The host galaxy is defined as the galaxy in the pair on which the bar structure analysis is performed.}, $\logm$ (x-axis), with three color-coded curves corresponding to subsamples divided by mass ratio, $\logmr$. Within each mass-ratio interval, the sample is further divided into two sub-bins according to the median value of $\logm$, resulting in two data points per mass-ratio subsample. This analysis is restricted to galaxy pairs with $\pdis < 25\kpc$ to focus on the most strongly interacting systems. 

We use two-sample proportion Z-tests to assess whether the bar fraction of paired galaxies is statistically significantly lower than that of control galaxies within each sub-bin, with the corresponding $p$-values listed in Table \ref{table1}. As shown in the figure, the dependence of $\Delta\fb$ on host galaxy properties varies with mass ratio. A statistically significant reduction in the bar fraction of paired galaxies is detected only in the major merger regime ($-0.5 < \logmr < 0.5$), and specifically in the higher-mass sub-bin of this interval ($p < 0.05$ in Table \ref{table1}). This indicates that, in major mergers, the bar fraction of paired galaxies decreases more strongly for more massive hosts. In contrast, $\Delta\fb$ values are close to zero and statistically indistinguishable from zero in all other sub-bins, including both minor-merger regimes ($-1.5 < \logmr < -0.5$ and $0.5 < \logmr < 1.5$), implying that the bar fraction of paired galaxies is comparable to that of isolated controls in these cases.

More generally, our results are consistent with previous studies of galaxy pairs, in which major merger pairs are found to produce the strongest tidal interactions to reshape the morphologies \citep[e.g.,][]{Cavanagh2020}. In such systems, especially when the member galaxy is massive, tidal perturbations are expected to be more violent and dynamically disruptive. These stronger tidal forces can more efficiently weaken pre-existing bars, leading to the observed reduction in bar fraction. This interpretation naturally explains why the mass-dependent response of bar structures is most clearly detected in major merger systems, while similar effects are not statistically evident in minor mergers.

\begin{table*}[htbp]
    \centering
    \caption{Two-Sample Z-Test Results of sub-samples with different galaxy properties}
    \label{table1}
        \begin{tabular}{c c c c c}
        \toprule
        Parameter & Mass Ratio Bin & Sub-Bin & Z-value & $p$-value \\
        \midrule
        \multirow{6}{*}{$\log M_\star$}
        & \multirow{2}{*}{$(-1.5, -0.5)$} & $[10.00, 10.31]$ & $-0.13$ & $0.90$ \\
        & & $[10.31, 11.50]$ & $-0.25$ & $0.80$ \\
        & \multirow{2}{*}{$(-0.5, 0.5)$} & $[10.00, 10.74]$ & $-0.44$ & $0.66$ \\
        & & $[10.74, 11.50]$ & $-4.39$ & $1.16\times10^{-4}$ \\
        & \multirow{2}{*}{$(0.5, 1.5)$} & $[10.00, 10.79]$ & $-1.66$ & $0.10$ \\
        & & $[10.79, 11.50]$ & $-0.93$ & $0.35$ \\
        \midrule
        \multirow{6}{*}{$B/T$}
        & \multirow{2}{*}{$(-1.5, -0.5)$} & $[0.00, 0.25]$ & $-0.70$ & $0.49$ \\
        & & $[0.25, 0.50]$ & $0.27$ & $0.79$ \\
        & \multirow{2}{*}{$(-0.5, 0.5)$} & $[0.00, 0.29]$ & $-2.20$ & $0.03$ \\
        & & $[0.29, 0.50]$ & $-2.80$ & $0.01$ \\
        & \multirow{2}{*}{$(0.5, 1.5)$} & $[0.00, 0.32]$ & $-2.17$ & $0.03$ \\
        & & $[0.32, 0.50]$ & $-0.54$ & $0.59$ \\
        \midrule
        \multirow{6}{*}{$A_i$}
        & \multirow{2}{*}{$(-1.5, -0.5)$} & $[0.00, 49.50]$ & $0.81$ & $0.42$ \\
        & & $[49.50, 90.00]$ & $-1.06$ & $0.29$ \\
        & \multirow{2}{*}{$(-0.5, 0.5)$} & $[0.00, 45.00]$ & $-2.57$ & $0.01$ \\
        & & $[45.00, 90.00]$ & $-2.20$ & $0.03$ \\
        & \multirow{2}{*}{$(0.5, 1.5)$} & $[0.00, 53.00]$ & $-1.40$ & $0.16$ \\
        & & $[53.00, 90.00]$ & $-1.24$ & $0.22$ \\
        \bottomrule
        \end{tabular}
\end{table*}

\subsection{Bulge-to-Total Ratio}

Bars are disk instabilities, and their formation is influenced by both internal structure and external perturbations. The bulge plays a key stabilizing role in this context. A more massive bulge increases the central gravitational potential, helping suppress bar formation by stabilizing the disk against global instabilities \citep[e.g.,][]{Athanassoula2003, Nair2010}. In addition, bulges can shield galaxies from external tidal forces during interactions, reducing the likelihood that a companion will destroy the structure of primary galaxies. To explore this, we use the bulge-to-total light ratio ($\BT$) to examine whether the stabilizing effect of the bulge modifies the bar suppression seen in galaxy pairs.

The middle panel of Figure~\ref{fig:df-mmr} presents the difference in bar fraction ($\Delta\fb$) between paired and control galaxies as a function of the bulge-to-total ratio $\BT$, with three color-coded curves corresponding to different mass-ratio bins ($\logmr$). Within each mass-ratio interval, the sample is divided into two sub-bins based on the median $\BT$. Two-sample proportion Z-tests reveal statistically significant reductions in bar fraction for paired galaxies in both $\BT$ sub-bins of major mergers ($-0.5 < \logmr < 0.5$) and in the low-$\BT$ sub-bin of minor mergers ($0.5 < \logmr < 1.5$), while no significant differences are detected in the remaining sub-bins ($p>0.05$). These results demonstrate that $\Delta\fb$ depends on $\BT$ in a mass-ratio dependent manner, with major mergers exhibiting bar suppression across all $\BT$ regimes, whereas minor mergers ($0.5 < \logmr < 1.5$) show suppression only for galaxies with smaller bulges.

It is worth noting that $\BT$ is correlated with stellar mass, as galaxies with more prominent bulges tend to be more massive. However, a comparison between the middle and top panels of Figure~\ref{fig:df-mmr} indicates that stellar mass and bulge size play partially independent roles. This distinction is particularly evident in the minor merger regime ($0.5 < \logmr < 1.5$), where a statistically significant suppression of the bar fraction is detected only for galaxies with small bulges. This behavior is consistent with a physical picture in which bulges can partially shield galaxies from external tidal perturbations during interactions, thereby reducing the likelihood that a companion disrupts the bar structure of the host galaxy \citep[e.g.][]{Cox2008}. Taken together, these results suggest that although bulge prominence plays a non-negligible role in protecting bar structures, stellar mass serves as a more primary parameter in regulating the response of bars to galaxy–galaxy interactions than bulge prominence alone.

\subsection{Relative Position Angle}

Numerical simulations have shown that the spin direction of galaxies can significantly influence the evolution of bars during interactions \citep{Lokas2018, Cavanagh2020, Kumar2022}. Since direct measurements of galaxy spin directions are challenging, the position angle (PA) of the galaxy's disk is commonly used as a proxy for the spin direction in observational studies. In this section, we investigate the relationship between the position angle and the bar fraction in galaxy pairs. 

We define the relative position angle as the angle between the position angles of the two galaxies in a pair, $A_i$. The position angle is measured clockwise from the $y$-axis of the image to the semi-major axis of the galaxy's disk, which is taken from \citep{Simard2011}. In this study, we explore how this relative position angle correlates with the bar fraction difference ($\Delta\fb$), shedding light on how tidal interactions at varying angles affect bar formation and disruption.

The bottom panel of Figure \ref{fig:df-mmr} presents $\Delta\fb$ as a function of the relative position angle ($A_i$), with three color-coded curves corresponding to different mass-ratio bins ($\logmr$). Following the same binning strategy as in the upper and middle panels, each mass-ratio interval is divided into two sub-bins based on the median value of $A_i$. Two-sample proportion Z-tests—conducted to assess whether the bar fraction of paired galaxies is significantly lower than that of control galaxies within each sub-bin. Table \ref{table1} reveals that statistically significant $\Delta\fb$ reductions are only detected in both $A_i$ sub-bins of the major merger regime. In contrast, $\Delta\fb$ values remain close to zero and statistically indistinguishable from zero across all $A_i$ sub-bins of the minor merger regimes.

These results indicate that the relative position angle does not have a statistically robust effect on the suppression or disruption of bar structures. Instead, the significant reduction in bar fraction is confined to the major-merger interval ($-0.5 < \logmr < 0.5$) and is primarily associated with the stellar mass of the host galaxy. This suggests that, for close pairs, the impact of interactions on bar evolution is governed mainly by mass-related factors rather than by the relative orientation of the interacting galaxies.

\section{Discussion}\label{discuss}

\subsection{Impact of Observational Effects}

In this section, we examine whether observational effects could influence our results on the bar fraction difference between paired and control galaxies.

First, we consider the impact of bar identification as a function of redshift. The degradation of spatial resolution with increasing redshift inevitably reduces the detectability of bars in more distant galaxies. However, because galaxies in the paired and control samples are matched in redshift, both samples are affected by resolution limitations in a similar manner. As a result, redshift-dependent resolution effects are expected to largely cancel out when comparing the two samples, and should not introduce a systematic bias in the bar fraction difference. In the extreme case where bars become entirely undetectable at high redshift, the measured bar fractions of both paired and isolated galaxies would approach zero, yielding a bar fraction difference close to zero. Therefore, resolution effects associated with redshift would tend to dilute the bar fraction difference rather than produce or enhance it. The observed lower bar fraction in paired galaxies is thus unlikely to be driven by resolution effects.

Second, we assess the potential impact of galaxy inclination on bar identification. Bars are increasingly difficult to detect in highly inclined systems, and our analysis is therefore restricted to low-inclination disk galaxies through an ellipticity selection. In interacting systems, tidal features may bias the photometric ellipticity measurement, causing disks to appear more elongated and leading to an overestimation of inclination. Under such a selection, paired galaxies that pass the ellipticity cut may in fact have lower true inclinations than control galaxies. This bias would make bars easier to detect in the paired sample, artificially increasing the measured bar fraction in pairs and biasing the bar fraction offset upward. Since we nevertheless observe a negative bar fraction offset, any inclination-related bias of this kind would act against our observed trend rather than produce it.

In summary, the observational effects considered here cannot account for the reduced bar fraction observed in close galaxy pairs. If anything, these effects would tend to weaken the observed difference. We therefore conclude that the decrease in bar fraction in close galaxy pairs reflects a genuine physical effect rather than an artifact of observational limitations or sample selection.

\subsection{Role of Tidal Interaction}

Our observations reveal a decrease in the bar fraction ($\fb$) among galaxy pairs with $\pdis < 25\kpc$, suggesting a possible connection between close encounters and bar suppression. However, a fraction of close pairs selected by projected separation are not truly interacting systems, but rather chance alignments along the line of sight \citep[e.g.,][]{Soares2007, Feng2020}. In genuinely interacting galaxies, tidal forces can drive pronounced morphological disturbances \citep[e.g., tidal tails][]{2010MNRAS.404..575L}. Therefore, the presence of such disturbance signatures provides an effective way to identify physically interacting systems within close-pair samples. To more robustly test whether the observed bar deficit is associated with tidal interactions, we examine how $\Delta\fb$ correlates with independent morphological indicators of interaction.

To quantify interaction-induced morphological disturbances, we adopt the automated interaction/merger classification from Galaxy Zoo DESI \citep{2023MNRAS.526.4768W}. This catalogue uses deep-learning models trained on Galaxy Zoo volunteer classifications from DESI Legacy Imaging Surveys \citep{2022MNRAS.509.3966W}, and provides posterior probabilities for the galaxy interaction decision tree with four outcomes: \emph{Merging}, \emph{Major Disturbance}, \emph{Minor Disturbance}, and \emph{None}. We use these classifications to separate close pairs with clear disturbance signatures from those that are likely projection-contaminated, and then reassess the bar fraction differences accordingly.

\begin{figure*}[ht!]
    \plotone{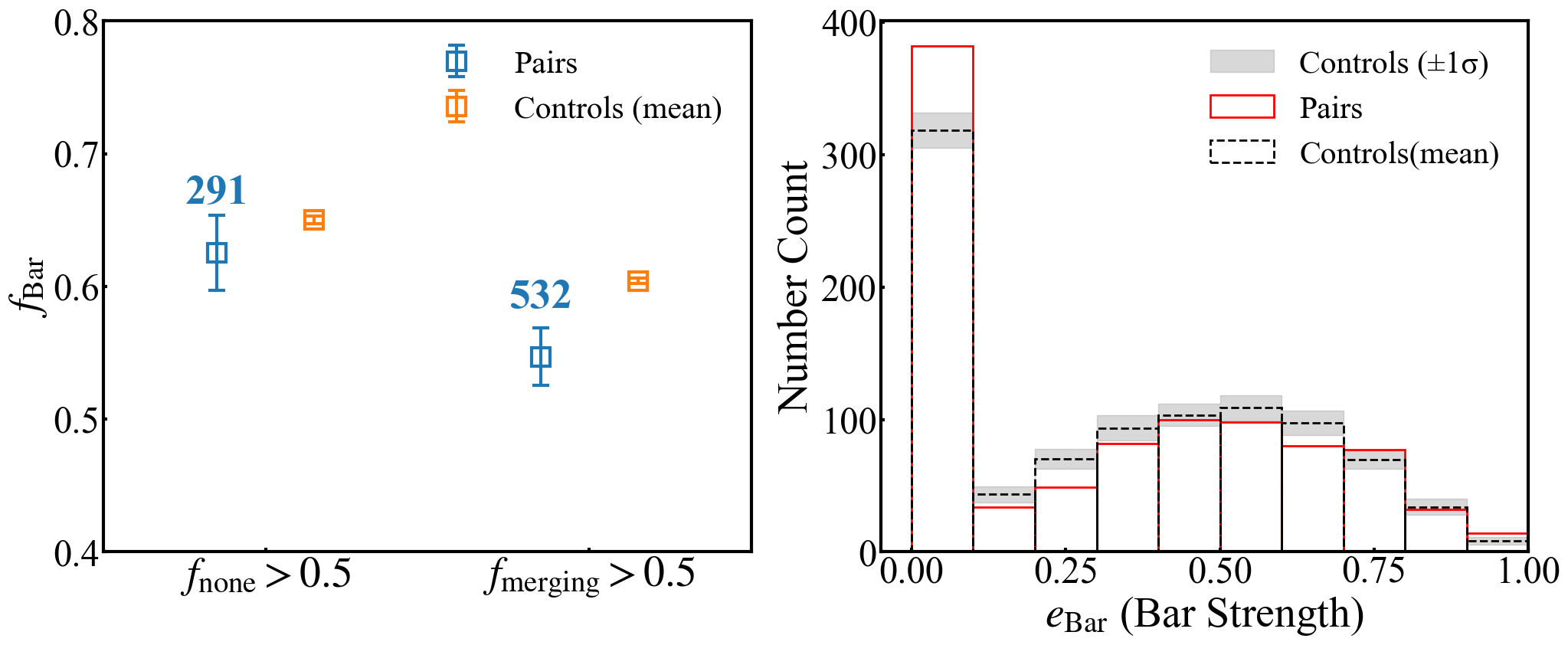}
    \caption{\textit{Left} panel: $\fb$ of close-pair subsamples classified by the presence or absence of merger/disturbance signatures within $\pdis < 25\kpc$. Blue and orange bars represent paired galaxies and their matched controls, respectively. Error bars indicate the uncertainties in $\fb$. The corresponding sample sizes are annotated for each subsample. \textit{Right} panel: Distribution of bar strength ($\eb$) for paired galaxies (red solid histogram) and control galaxies. Here, the black dashed histogram denotes the mean of 100 independent control samples, while the gray shaded region represents the $\pm 1\sigma$ scatter across these 100 iterations. The $y$-axis shows raw galaxy counts}
    \label{fig:df-shape}
\end{figure*}

Based on this classification, we divide the close pairs with $\pdis<25\kpc$ into two subsamples: a \textsc{merging} subsample defined by $f_{\rm merging} > 0.5$, where $f_{\rm merging}$ is the sum of the \emph{Merging}, \emph{Major Disturbance}, and \emph{Minor Disturbance} probabilities, and a \emph{non-merging} subsample defined by $f_{\rm none} > 0.5$, where $f_{\rm none}$ corresponds to the probability of \emph{None} in the Galaxy Zoo classification. For each close-pair subsample, we compare its bar fraction with that of its correspondingly matched control galaxies. The left panel of Figure \ref{fig:df-shape} shows the resulting $\fb$ measurements for the two close-pair subsamples and their controls, sample sizes are annotated for paired samples. Two-sample proportion Z-tests are used to assess the statistical significance of the bar fraction differences between paired galaxies and their matched controls, with the corresponding $p$-values shown in Table \ref{table2}.

\begin{table}[htbp]
    \centering
    \caption{Two-sample proportion Z-tests for Interaction and Weak/Strong Bar Grouping}\label{table2}
    \begin{tabular}{l c c}
    \toprule
    Category & Z-value & $p$-value \\
    \midrule
    \emph{None} & $-0.88$ & $0.38$ \\
    \emph{Merging} & $-2.68$ & $0.01$ \\
    \midrule
    Weak Bar & $-4.18$ & $2.94\times10^{-4}$ \\
    Strong Bar & $1.09$ & $0.27$ \\
    \bottomrule
    \end{tabular}
\end{table}

The results demonstrate that merging galaxies exhibit a significantly lower bar fraction than their control counterparts ($p = 0.01$), whereas the bar fraction of non-merging galaxies shows only a mild reduction that is not statistically significant ($p > 0.05$). This indicates that the observed decrease in bar fraction in close galaxy pairs is primarily associated with systems undergoing merging or strong morphological disturbance. In other words, bar suppression is largely driven by galaxies experiencing intense tidal interactions, which are more effective at disrupting existing bars.

\subsection{The Distribution of Bar Strengths}

The previous sections have focused on the overall bar fraction. Although informative, the bar fraction alone cannot fully capture the complexity of bar evolution under the influence of tidal interactions. To gain a more complete picture, it is essential to display how the strength of bars is affected in galaxy pairs.

The right panel of Figure \ref{fig:df-shape} shows the distribution of bar strength ($\eb$) for paired (red solid histogram) and control (black dashed histogram) galaxies within the projected separation range of $\pdis < 25\kpc$ (the $y$-axis denotes raw galaxy counts. Visually, a prominent excess is observed in the counts of paired galaxies with $\eb < 0.1$ compared to the control sample, indicating that a significantly larger fraction of galaxy pairs exhibit either extremely weak bars or no detectable bar structure. Conversely, in higher $\eb$ bins, the count of paired galaxies falls below that of the controls. This deficit is particularly evident in the ranges $0.2 < \eb < 0.3$ and $0.6 < \eb < 0.7$, where paired galaxies show a noticeable suppression relative to the control population. At $\eb > 0.7$, however, the distributions of paired and control galaxies become comparable.

To quantify the statistical significance of bar strength differences, we perform Two-sample proportion Z-tests on the fractions of weak bars ($\eb < 0.7$) and strong bars ($\eb > 0.7$), and $p$ values are shown in Table \ref{table2}. The results show that the fraction of weak bars in paired galaxies is significantly lower than that in the control sample ($p< 0.05$), whereas the fraction of strong bars exhibits no statistically significant difference between the two samples ($p > 0.05$). To assess the robustness of this conclusion, we further test bar strength thresholds ranging from $0.6$ to $0.8$ for distinguishing weak and strong bars. No substantial change in the statistical significance is observed across these thresholds, confirming that the suppression of weak bars in paired galaxies and the insensitivity of strong bars to interactions are robust to the choice of bar strength boundary.

These results imply that the observed decrease in bar fraction among galaxy pairs is primarily driven by the destruction or weakening of weak bars ($\eb < 0.7$), which are more susceptible to tidal perturbations. In contrast, strong bars ($\eb > 0.7$) appear to be more resilient against the disruptive effects of close companions.

\subsection{Trigger vs Suppress}

Our observations reveal a lower bar fraction in galaxy pairs within $ \pdis < 25$ kpc, which we interpret as evidence that galaxy interactions can destroy pre-existing bars. However, this does not preclude the possibility that interactions can also trigger bar formation. Both effects may coexist, operating on different timescales and under different conditions.

Numerical simulations have shown that interaction-induced bars often exhibit a delayed response: bar formation does not occur immediately at the time of pericentric passage but rather develops as the galaxies begin to separate after the closest encounter \citep[e.g.,][]{Peschken2019}. In contrast, tidal disruption to galactic structure peaks near pericenter, when gravitational forces are strongest \citep[e.g.,][]{Lotz2008}.

Therefore, in close pairs, typically dominated by systems that have just undergone the pericentric passage, destructive processes may dominate over bar triggering, leading to a net decrease in bar fraction. In contrast, wide pairs ($\pdis > 25\kpc$) have either not yet interacted or have already passed through pericenter. Although these pairs include a non-negligible fraction of physically interacting systems \citep{Patton2013, Feng2020}, we do not observe a similar reduction in bar fraction. This could indicate that bars previously weakened by interactions may have reformed during the separation phase \citep{Zana2018}, or that the destructive and constructive influences of interactions reach a balance at these separations. To disentangle these competing processes and better understand the net impact of interactions on bar evolution, future studies should follow bar properties along the merger sequence.

\section{Summary}\label{sec:sum}

We investigate the impact of galaxy interactions on bar fractions using a sample of $20{,}264$ galaxies in galaxy pairs selected from the SDSS, with bar identifications adopted from the catalog of \citet{Deng2023} based on elliptical isophote fitting. To isolate interaction-driven effects, we construct a carefully matched control sample of isolated galaxies, matched in redshift, stellar mass, bulge-to-total ratio, and local neighbor density.

We find that galaxy interactions lead to a significant suppression of bars in close pairs. Specifically, paired galaxies exhibit a clear reduction in bar fraction, with $\Delta\fb \approx -0.06$ at projected separations $\pdis < 25\kpc$. This reduction is almost entirely driven by systems exhibiting clear merger or disturbance signatures, indicating that the observed bar deficit is closely associated with the disruptive effects of tidal perturbations during galaxy interactions.

We further explored how the bar suppression signal relates to galaxy-pair properties. The primary driver of significant bar suppression is the merger type. Robust and widespread bar suppression is only detected in the major-merger regime ($-0.5 < \logmr < 0.5$). In contrast, minor-merger regimes generally exhibit negligible bar suppression, with the sole exception of more massive galaxies hosting small bulges. The relative position angle $A_i$ shows no significant correlation with $\Delta\fb$ across any regime.

Finally, the observed decrease in bar fraction in close galaxy pairs is primarily driven by the suppression of weak bars ($\eb < 0.7$), while strong bars ($\eb > 0.7$) show no significant difference between paired and control galaxies. This indicates that weak bars are more vulnerable to tidal perturbations, whereas strong bars are largely resilient to close interactions.

\begin{acknowledgments}
We thank the anonymous referee for constructive comments that greatly helped to improve this paper. We thank Min Du for the useful discussions. This study is supported by the National Natural Science Foundation of China (Nos. 11903012, 12103017, 12141302, 12173013), Natural Science Foundation of Hebei Province (No. A2025205037), and China Manned Space Project (Nos. CMS-CSST-2021-A09, CMS-CSST-2025-A07). LL acknowledge the financial support from the Physics Postdoctoral Research Station at Hebei Normal University. SS thanks support from National Key R\&D Program of China (No. 2025YFF0510603).

\end{acknowledgments}

\bibliography{ref.bib}
\bibliographystyle{aasjournalv7}

\end{CJK*}
\end{document}